\documentclass[a4paper,11pt]{article}
\pdfoutput=1 

\usepackage{jinstpub} 

\title{Design of a novel compact neutron collimator}

\author[a]{R. Bedogni,}
\author[b,c]{M. Costa,}
\author[b,c]{E. Durisi,}
\author[d,e]{F. Grazzi,}
\author[a,b]{A. Lega,}
\author[b,c]{E. Mafucci,}
\author[a,b]{L. Menzio,}
\author[b,c]{V. Monti,}
\author[b,c,1]{O. Sans Planell,\note{Corresponding author.}}
\author[b,c]{L. Visca,} 

\affiliation[a]{INFN, Laboratori Nazionali di Frascati,\\Via Enrico Fermi 40, 00044, Frascati, Italy}
\affiliation[b]{Universita degli Studi di Torino,\\Via Pietro Giuria 1, 10125, Torino, Italy}
\affiliation[c]{INFN, Sezione di Torino,\\Via Pietro Giuria 1, 10125, Torino, Italy}
\affiliation[d]{Consiglio Nazionale delle Ricerche, Istituto dei Sistemi Complessi,\\ Via Madonna del Piano 10, I-50019 Sesto Fiorentino, Italy}
\affiliation[e]{INFN, Sezione di Firenze,\\Via Sansone 1, 50019, Sesto Fiorentino, Italy}
\emailAdd{oriol.sansplanell@to.infn.it}

\abstract{In this work the concept of a novel slow neutron collimator and the way to operate it are presented. The idea is based on the possibility to decouple the device's field-of-view from its collimation power. A multi-channel geometry is proposed consisting of a chess-board structure where highly neutron-absorbing channels are alternated to air channels. A borated polymer was purposely developed to produce the attenuating components in the form of square-sectioned long rods. A scalable structure consisting of multiple collimation sectors can be arranged. The geometrical parameter L/D, corresponding to the ratio between the length of a channel and its width, defines the collimation power. Several sectors can be arranged one after the other to reach relevant collimation powers. Each sector, 100 $mm$ long, is composed by several channels with D = 2.5 $mm$ corresponding to an L/D coefficient of 40. The target field of view is 50x50 $mm^2$. This novel collimator, developed inside the INFN-ANET collaboration, due to its intrinsic compactness, will be of great importance to enhance the neutron imaging capability of small to medium-size neutron sources.}


\arxivnumber{1234.56789} 

\begin{document}
\maketitle
\flushbottom

\section{Introduction}

Neutron collimators found in imaging or scattering experiments are mostly classical long-tube, pin-hole,  Soller slit, multi-plate, or honeycomb type collimators. In each case, the objective of the collimator is to efficiently reduce the dispersion angle of the neutron beam. The main factor to determine the performance of a collimator is the ratio between the transmission length of the channel (L) and its width (D). Typical L/D values are in the range 100-1000 (e.g. \cite{h}). \newline
The ANET - Advanced NEutron Techniques - project aims at creating a compact thermal neutron collimator (CNC) for imaging applications with a length below 1 $m$ and an L/D factor greater than 100. This would allow small or medium-size facilities, such as small reactors or accelerator based neutron sources, to access neutron imaging applications. The core idea behind this design is to create a multi-channel, scalable structure capable of delivering a highly collimated thermal neutron beam within a short distance. The design consists of an alternate sequence of air and highly absorbant channels arranged in a chessboard section geometry \cite{e}. The absorbing rods are manufactured using a borated polymer purposely created to achieve satisfactory nuclear and mechanical performances. The use of air as transmission channel medium is not limiting the collimation power and it results in an overall simplification of the collimation structure.

\section{State-of-the-art}
By the early XXI century, multi-channel neutron collimators were already a common technology, as proven by the article by Petrillo et al. \cite{a} with their proposal of a gadolinium and aluminium honeycomb collimator. More recently, Stone et al. \cite{b} demonstrated the possibility of boron carbide and plastic compounds as effective materials for the building of neutron collimators. There are extensive studies regarding microchannel neutron collimators \cite{c,d} reaching up to thousands of channels per square millimetre, but they suffer the drawback of a relatively high transmission index for uncollimated neutrons.\newline
In our approach, the designed collimator will implement a matrix of millimetre size transmission channels alternated to absorbing elements made of a mixture of polylactide and natural boron carbide, which  will offer a very low transmission index for uncollimated neutrons.

\section{Design of the ANET compact neutron collimator}

\begin{figure}[h]
\centering 
\includegraphics[width=0.9\textwidth]{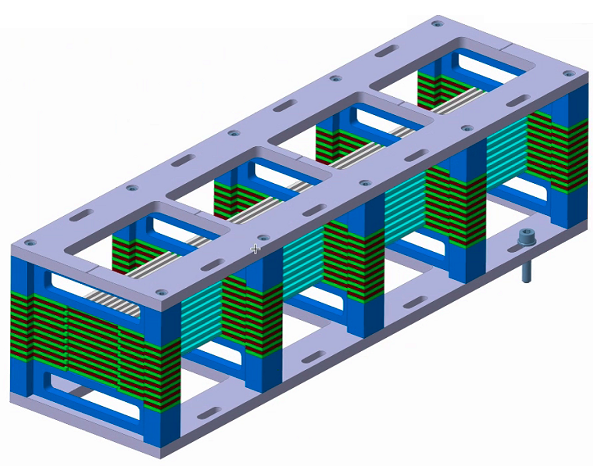}
\caption{\label{fig:collimator-render2} 3D rendering of the ANET compact neutron collimator.}
\end{figure}

\begin{figure}[h]
\centering 
\includegraphics[width=0.9\textwidth]{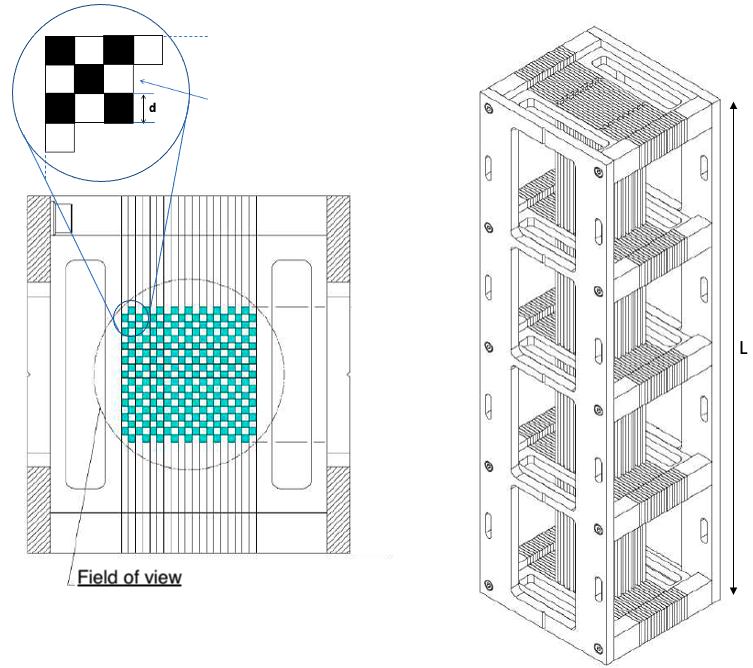}
\caption{\label{fig:collimator-design} Design of the collimator structure. D is the dimension of a single channel and L the length of the collimator structure.}
\end{figure}

\begin{figure}[h]
\centering 
\includegraphics[width=0.9\textwidth]{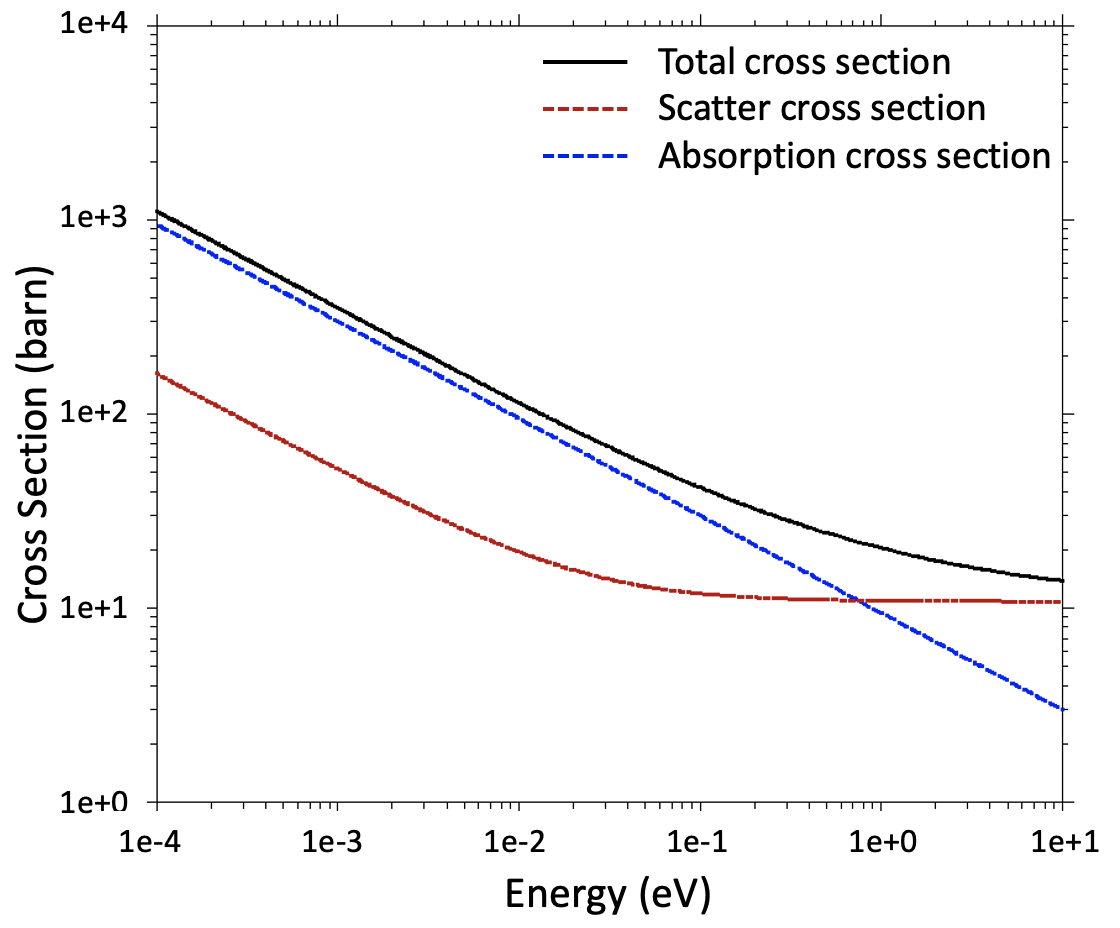}
\caption{\label{fig:total-XS-material} Total, scatter and absorption cross-section of the material used to build the collimator absorbing rods as a function of the neutron energy.}
\end{figure}

The idea is to decouple the field of view and the collimation power of the instrument. It consists of a sequence of air and highly absorbent material channels organized in a chessboard-like section geometry. In order to uniformly illuminate the object under study the instrument will be mounted on a moving stage system. Figure \ref{fig:collimator-render2} shows a 3-D rendering of the collimator. \newline
The collimator consists of four stages arranged inside an aluminium structure to reach the required mechanical rigidity and maintain a low structural activation. Each stage consists of 100 absorbing rods, each one 100 $mm$ in length and with a section of 2.5 x 2.5 $mm^2$, aligned using laser-cut thin comb-like plates. The expected L/D ratio is 160. The present collimator has been designed to cover an area of 50 x 50 $mm^2$. Figure \ref{fig:collimator-design} shows the technical design of the collimator structure. It is possible to increase the total field of view of the collimator or its length just by including more singular elements on a stage or increasing the number of stages itself. The ANET design is scalable both longitudinally and transversely, reaching elevated collimation factors within a limited space.
The material of the absorbing rods is a combination of polylactide and boron carbide. Figure \ref{fig:total-XS-material} shows the cross-sections (total, absorption and scattering neutron cross-sections respectively) on the material extracted from the simulations performed with MCNP6 transport code \cite{g} using ENDF/B VII.1 cross-section libraries \cite{f}.
From the figure, it is clear that the total cross-section in the thermal energy range (below $10^{-2} eV$) is of the order of $10^{2}$ $barn$ and mostly due to absorption. This high absorption cross-section makes the material an ideal candidate for building the absorbing elements of the compact neutron collimator. \newline
\section{Material performance study}
\begin{figure}[h]
\centering 
\includegraphics[width=1\textwidth]{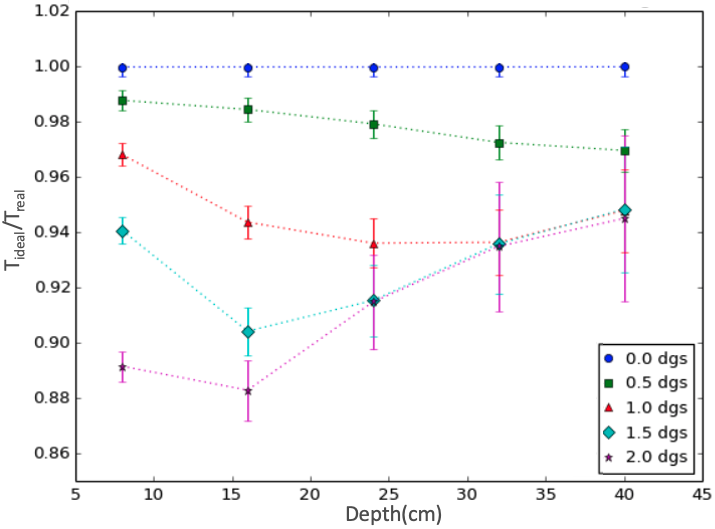}
\caption{\label{fig:ratio-ideal-real} Distribution of the ratio $T_{ideal} / T_{real}$ as a function of the collimator depth. The curves correspond to different choices of the beam divergence as quoted in the legend.}
\end{figure}

The evaluation of the attenuation and scattering effects induced by the collimator rods, taking into account their composition with respect to an ideal attenuator has been evaluated using MCNP6.\newline
To calculate the mean free path of a neutron through the material the following equation \ref{eq:macroscopic-XS} is used:
\begin{equation}
    \Sigma_T = \sigma_T N
    \label{eq:macroscopic-XS}
\end{equation}
where $\Sigma_T$ is the total macroscopic cross-section of the material, $\sigma_T$ the total microscopic cross-section, $N$ atomic number density, expressed in nuclei over $cm^3$. Assuming a microscopic total cross section at thermal energies of 100 $barn$ (see figure \ref{fig:total-XS-material}), the macroscopic cross-section is 1.19 $cm^{-1}$. Since the mean free path $\lambda$ of a particle is the inverse of the macroscopic cross-section, this results in a value $\lambda = 8.4 mm$. Each rod of the ANET collimator introduces on its longitudinal axis an absorbing factor of $2\cdot10^4$. The prototype is expected to have 4 sectors, with a total length of 400 $mm$, thus providing a very high selection of neutrons with the desired direction. The material with the appropriate multi-channel geometry has been simulated using MCNP6 assuming a 400$mm$ long collimator. In the simulation a monoenergetic thermal source has been assumed with selectable divergences. 
In order to quantify the material performance the comparison between the number of transmitted neutrons at a certain depth passing through a 50 x 50 $mm^2$ surface in the case of a multi-channel collimator composed by an ideal absorber ($T_{ideal}$) and by the real material ($T_{real}$) has been evaluated. The study has been done specifically at 8$cm$, 16$cm$, 24$cm$, 32$cm$ and 40$cm$ depths and under 5 different source divergences, from 0 degrees up to 2 degrees. The distributions of the ratio $\frac{T_{ideal}}{T_{real}}$ are shown in figure \ref{fig:ratio-ideal-real}.
The ratio $\frac{T_{ideal}}{T_{real}}$ shall always be  smaller than 1, as the ideal collimator will always absorb more neutrons than the real one. An asymptotic behaviour of $\frac{T_{ideal}}{T_{real}}$ towards 1 with increasing depth is expected. At the nominal depth of 40 $cm$ the ratio is above 0.95 for all the initial divergence scenarios. Figure \ref{fig:ratio-ideal-real} also shows that the effect of neutrons laterally exiting their initial channel tends to be minimised with the increase of the collimator's length.

\section{CNC method of operation}
\begin{figure}[h]
\centering 
\includegraphics[width=0.9\textwidth]{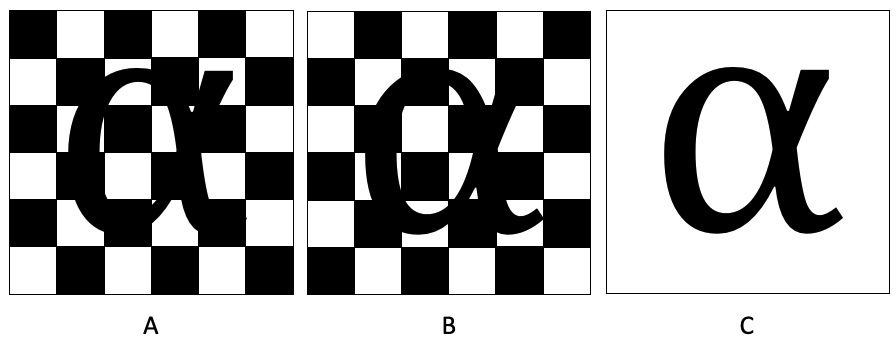}
\caption{\label{fig:moving_collimator} Schematic design of the resulting image combining two collimator positions on a fixed object. Images A and B emulate a neutron radiography with the CNC in two complementary positions, and C the result of combining the two former images.}
\end{figure}

One apparent disadvantage of the chessboard-like design of the compact neutron collimator is the alternate structure of light and dark areas due to the geometry. In order to have each part of an object uniformly illuminated, the instrument needs to be moved with respect to the object. At least two images taken with a translation equivalent to a channel width from one another are needed. Figure \ref{fig:moving_collimator} highlights this concept, in which the combination of two images with the collimator placed in compatible positions eliminates the chessboard pattern as a result. More sophisticated techniques will be studied in order to optimise the quality and resolution of the neutron radiographies, minimising the systematic artefacts caused by the structure.

\section{Conclusions and future work}
The concept of an innovative compact collimator for thermal neutron radiography has been presented. It will significantly benefit neutron imaging applications, due to the combination of the following effects:
\begin{enumerate}
    \item Reduction of the design constraints for neutron science facilities.
    \item Reduction of the primary beam attenuation.
    \item Elimination of the need for filling gas or to operate it under vacuum. 
\end{enumerate}
This will open the possibility for a broader array of facilities, such as small and medium-sized reactors and accelerator-based facilities, to access neutron imaging technique, with a sensitive reduction of costs and a wider application range.\newline
The work performed so far confirms promising perspectives for this collimator concept. An intense measuring campaign is foreseen in the near future, to demonstrate how to practically operate the ANET CNC in order to quantify its real performance parameters and to setup and study methods for optimising the image results.  

\acknowledgments
The project is funded by the Italian National Institute of Nuclear Physics (INFN) under the National Scientific Comission V for "Technological Research and Development".


\end{document}